\documentclass[twocolumn,aps,pre,amsmath,amssymb]{revtex4-1}
\usepackage{graphicx}
\usepackage{dcolumn}
\usepackage{bm}
\usepackage{multirow}
\usepackage{amssymb}
\usepackage{amsmath}
\usepackage{graphicx}
\usepackage{xcolor}
\usepackage[colorlinks,citecolor=blue,linkcolor=red,urlcolor=blue]{hyperref}
\usepackage{CJK}
\usepackage{indentfirst}
\usepackage{amsmath}
\usepackage{cases}

\begin{document}
\title{Quantum criticality in the disordered Aubry-Andr\'{e} model}
\author{Xuan Bu$^{1}$}
\author{Liang-Jun Zhai$^{2,3}$}
\author{Shuai Yin$^{1}$ }
\email{yinsh6@mail.sysu.edu.cn}
\affiliation{$^1$School of Physics, Sun Yat-Sen University, Guangzhou 510275, China}
\affiliation{$^2$Department of Physics, Nanjing University, Nanjing 210093, China}
\affiliation{$^3$The school of mathematics and physics, Jiangsu University of Technology, Changzhou 213001, China}

\date{\today}
\begin{abstract}
In this paper, we explore quantum criticality in the disordered Aubry-Andr\'{e} (AA) model. For the pure AA model, it is well-known that it hosts a critical point separating an extended phase and a localized insulator phase by tuning the strength of the quasiperiodic potential. Here we unearth that the disorder strength $\Delta$ contributes an independent relevant direction near the critical point of the AA model. Our scaling analyses show that the localization length $\xi$ scales with $\Delta$ as $\xi\propto\Delta^{-\nu_\Delta}$ with $\nu_\Delta$ a new critical exponent, which is estimated to be $\nu_\Delta\approx0.46$. This value is remarkably different from the counterparts for both the pure AA model and the Anderson model. Moreover, rich critical phenomena are discovered in the critical region spanned by the quasiperiodic and the disordered potentials. In particular, in the extended phase side, we show that the scaling theory satisfy a hybrid scaling form as a result of the overlap between the critical regions of the AA model and the Anderson localization.
\end{abstract}
\maketitle

\section{Introduction}
Quasiperiodic systems have recently gained growing attentions since a lot of novel phases and phase transitions emerge therein~\cite{Harper1955,Aubry1980,Sokoloff1981,Sarma1988,Biddle2009,Biddle2011,Lang2012,verbin2013,Zhang2018,
Ganeshan2013,Luschen2018,Skipetrov2018,Agrawal2020,Longhi2019,HepengYao2019,zhai2020,Goblot2020,zhai2020,YFWang2022}.
For example, unconventional superconductivity appears in the twisted-bilayer systems with incommensurate Moir\'{e} superlattices~\cite{CYuan2018}, non-trivial topological property comes up in one-dimensional ($1$D) quasicrystals owing to their profound connections to the topological insulators in higher dimensions~\cite{Kraus2012},
non-Fermi-liquid behavior arises in the strongly correlated quasicrystals~\cite{Andrade2015}, and so on.
Among various quasiperiodic systems, the Aubry-Andr\'{e} (AA) model stands out with many unusual characteristic features~\cite{Aubry1980,Harper1955}, such as the self-similar energy spectra and non-trivial topological properties.
In particular, the AA model features a remarkable self-dual extended phase-localizaiton transition at a multifractal critical point~\cite{Aulbach2004}.
The AA model has been realized in several experimental setups including ultracold atoms in bichromatic laser potentials~\cite{Billy2008} and photonic devices~\cite{Roati2008}.
Besides, a variety of generalized AA models are proposed and show rich localization properies~\cite{Sarma1988,Liu2015,Thouless1983},
such as the appearance of the mobility edge by including other hopping terms~\cite{Biddle2009,Biddle2011}, many-body localization by including the interaction~\cite{Wang2021,Strkalj2021,Xu2019,Mastropietro2015,Khemani2017,zhangsx2018,Zhang2019arxiv,YFWang2022},
and the open AA model with dissipation~\cite{Hamizaki2019,zhai2020,Jiang2019,Longhi2019,Liu20212,Schiffer2021,Zhai2022}.

On the other hand, disorder is ubiquitous in nature and has dramatic effects on static and dynamic properties in both classical and quantum systems~\cite{Evers2008}.
For instance, localization induced by disorder, proposed by Anderson, is a longstanding research topic in condensed matter physics~\cite{Anderson1958,Thouless1974,Yang2021,Pu2022}.
Theoretically, universality classes of Anderson transition have been categorized~\cite{Evers2008,Shinobu1981,Alexander1997,TWang2021,XLuo2021,XLuo2022}.
Experimentally, the Anderson localization has been realized in cold atomic gases~\cite{Semeghini2015,DHWhite2020}, quantum optics~\cite{Wiersma1997,Mookherjea2014}, acoustic waves~\cite{HFHu2008}, and electronic systems~\cite{Lee1985}.
In addition, understanding the effects of quenched disorder on continuous phase transitions is a question of enduring interest~\cite{Subirbook,MatthiasVojta2003}.
In periodic lattice systems, the divergence of the correlation length at the critical point gives universal critical phenomena that are controlled by renormalization-group fixed points of a translationally invariant continuum quantum field theory.
A quenched disorder configuration breaks translation symmetry in all length scales, and thus can produce behavior qualitatively different from that of a periodic systems~\cite{Subirbook,MatthiasVojta2003,Coleman2005}.
However, the disorder effects in the phase transitions in quasiperiodic systems are still largely unknown except for some recent works in many-body localization transitions~\cite{Khemani2017,zhangsx2018,Zhang2019arxiv}.

In this paper, we investigate the phase transition properties in the $1$D disordered AA model. The phase diagram is sketched in Fig.~\ref{phase}, in which $\Delta$ denotes the disorder strength and $\delta$ denotes the distance to the AA critical point in the direction of the quasiperiodic potential.
We find that the disorder strength $\Delta$ represents an independent relevant direction at the critical point of the AA model.
Our scaling analyses show that the localization length $\xi$ scales with $\Delta$ as $\xi\propto\Delta^{-\nu_\Delta}$ with $\nu_\Delta$ a new critical exponent, which is estimated to be $\nu_\Delta\approx0.46$.
This exponent is remarkably different from both the counterpart exponent $\nu_\delta$ by tuning the strength of the quasiperiodic potential~\cite{Sinha2019,Wei2019} and the counterpart one $\nu_A$ in the Anderson model without quasiperiodic potential~\cite{Wei2019}.
Furthermore, critical behaviors of the localization length $\xi$, the inverse participation ratio (IPR), and the energy gap $\Delta E$ in the critical region spanned by $\delta$ and $\Delta$ are studied.
A full scaling form is then developed to describe these critical properties. In particular, in the extended phase side with $\delta<0$, we show that the scaling behaviors are controlled by both the AA critical point and the Anderson transition.
Accordingly, the scaling theory satisfies a hybrid scaling form as a result of the overlap between the critical regions of the AA model and the Anderson localization.

The rest of the paper is arranged as follows.
After introducing the disordered AA model in Sec.~\ref{secmodel}, we focus on the scaling dimension of the disorder strength and determine the exponent $\nu_\Delta$ in Sec.~\ref{disorderd}.
Then in Sec.~\ref{scalingcr}, the critical properties in the critical region in the presence of the disorder is studied.
A summary is given in Sec.~\ref{secSum}.

\section{\label{secmodel}Model and characteristic observables}
The Hamiltonian of the disordered AA model reads
\begin{eqnarray}
\label{Eq:model}
H &=& -J\sum_{j}^{L}{(c_{j}^\dagger c_{j+1}+h.c.})\\ \nonumber
&&+(2J+\delta)\sum_{j}^{L}\cos{[2\pi(\gamma j+\phi)]c_j^\dagger c_j}\\ \nonumber
&&+\Delta \sum_{j}^{L}w_j c_j^\dagger c_j,
\end{eqnarray}
in which $c_j^\dagger (c_j)$ is the creation (annihilation) operator of the hard-core boson, $J$ is the hopping coefficient which is set as unity of the energy scale, $(2J+\delta)$ measures the amplitude of the quasi-periodic potential, $\gamma$ is an irrational number, $\phi\in[0,1)$ is phase of the potential, $w_j\in[-1,1]$ gives the quenched disorder configuration, and $\Delta$ measures the disorder strength.
The periodic boundary condition is imposed in the following calculation.
To satisfy the periodic boundary condition, $\gamma$ has to be approximated by a rational number $F_n/F_{n+1}$ where $F_{n+1}=L$ and $F_{n}$ are the Fibonacci numbers~\cite{Jiang2019,Zhai2022}.
Without the last disorder term, i.e., $\Delta=0$, it was shown that all the eigenstates of model~(\ref{Eq:model}) are localized when $\delta>0$, while all the eigenstates are delocalized when $\delta<0$~\cite{Aubry1980}.
In contrast, without the quasiperiodic potential, i.e., $\delta=-2J$, the system is always in the localized phase for any finite $\Delta$, indicating the Anderson localization transition point is at $\Delta=0$~\cite{Anderson1958,Lee1985}.

To describe the critical properties of the localization transition, some characteristic quantities are employed. First, in the localized phase, the localization length, $\xi$, is defined in the localization phase as~\cite{Sinha2019}
\begin{equation}
\label{Eq:xiscaling}
   \xi = \sqrt{\sum_{n>n_c}^{L} [( n - n_c )^2 ] P_i},
\end{equation}
in which $P_i$ is the probability of the wavefunction at site $i$, and $n_c\equiv\sum nP_i$ is the localization center.
Near a critical point, $\xi$ scales with the distance to the critical point $g$ as
\begin{equation}
\label{Eq:xiscaling1}
\xi\propto g^{-\nu}.
\end{equation}
For the pure AA model, $\Delta=0$, $g=\delta$, and $\xi\propto \delta^{-\nu_\delta}$ with $\nu_\delta=1$~\cite{Sinha2019,Wei2019}.
For the pure Anderson model, $\delta=-2J$, $g=\Delta$, and $\xi\propto \Delta^{-\nu_A}$ with $\nu_A=2/3$~\cite{Wei2019}.

The second quantity to characterize the localization transition is the inverse participation ratio (IPR), which is defined as~\cite{Bauer1990,Fyodorov1992}
\begin{equation}
\label{Eq:ipr}
{\rm IPR} = \frac{\sum_{j=1}^L||\Psi(j)\rangle|^4}{\sum_{j=1}^L||\Psi(j)\rangle|^2},
\end{equation}
where $|\Psi(j)\rangle$ is the eigenvector. For the metallic phase, the wave function is homogeneously distributed through all sites, and ${\rm IPR}$ scales as ${\rm IPR}\propto L^{-1}$, while for the localization state ${\rm IPR}\propto L^0$. At the localization transition point, ${\rm IPR}$ satisfies a scaling relation ~\cite{Zhai2022}
\begin{equation}
\label{Eq:iprscaling1}
{\rm IPR}\propto L^{-s/\nu}.
\end{equation}
When $L\rightarrow\infty$, ${\rm IPR}$ scales with $g$ as
\begin{equation}
\label{Eq:iprscaling2}
{\rm IPR}\propto g^{s}.
\end{equation}
For the pure AA model, $s_\delta\approx0.33$, while for the pure Anderson model, $s_A=2/3$, since its critical point is just the homogeneous extended state without disorder.

As in usual quantum criticality, the energy gap between the first excited state and the ground state can also be used to characterize the localization transition. According to the finite-size scaling, the energy gap $\Delta E$ should scales as
\begin{equation}
\label{Eq:gapscaling}
   \Delta E \propto L^{-z},
\end{equation}
for $g=0$. When $L\rightarrow\infty$, $\Delta E$ scales with $g$ as
\begin{equation}
\label{Eq:gapscaling2}
{\Delta E}\propto g^{-\nu z}.
\end{equation}
For the pure AA model, $z_\delta\approx2.37$~\cite{Cestari2011,Sinha2019,Wei2019}, while for the pure Anderson model, $z_A=2$~\cite{Wei2019}.

\begin{figure}[tbp]
\centering
  \includegraphics[width=0.8\linewidth,clip]{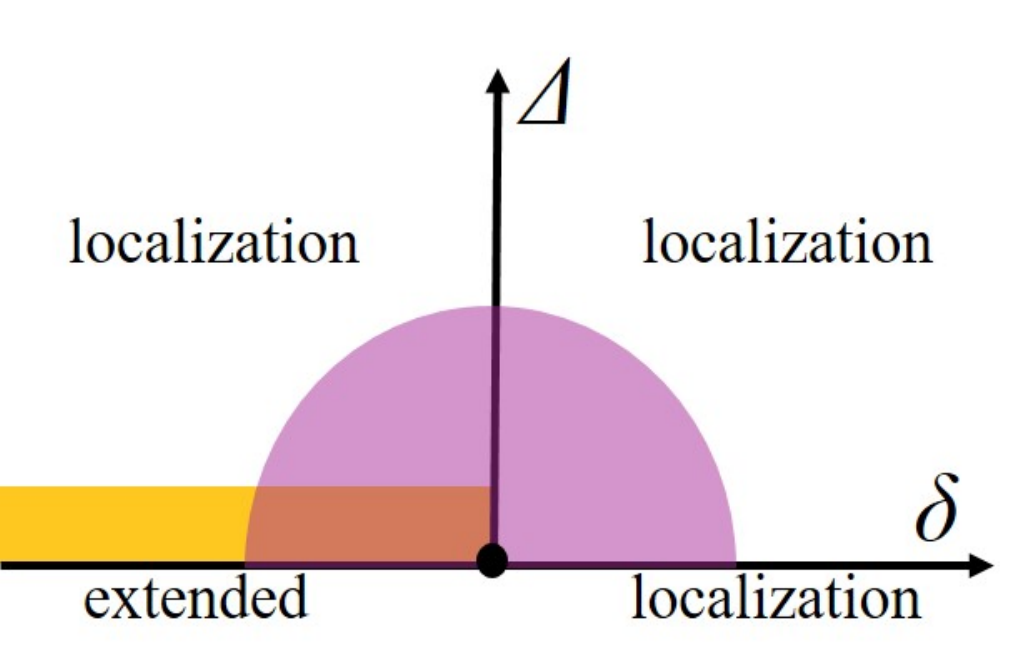}
  \vskip-3mm
  \caption{Sketch of the phase diagram of the disorder AA model. The violet region denotes the critical region of localization transition of the disordered AA model.
  The yellow region denotes the critical region of the Anderson localization transition.
  Near the critical point of $\delta=0$ and $\Delta=0$, these critical regions overlap with each other.
  }
  \label{phase}
\end{figure}

Since in $1$D non-interacting systems can be localized even for infinitesimal disorder~\cite{Anderson1958,Evers2008,Lee1985}, it is expected that the disordered AA model~(\ref{Eq:model}) is always in the localized phase except for the axis with $\Delta=0$ and $\delta<0$, as show in Fig.~\ref{phase}. For $\delta<0$ and small $\Delta$, the system is in the critical region of the Anderson localization, as illustrated by the yellow region in Fig.~\ref{phase}. However, when $\delta=0$, it is expected that the critical dimension of $\Delta$ should be different from the one for $\delta<0$. Moreover, a critical region is spanned by $\Delta$ and $\delta$ near the AA critical point, as illustrated by the purple region in Fig.~\ref{phase}. In the following, we will determine the dimension of $\Delta$ and investigate the critical properties in the critical region surrounding the AA critical point.

\section{\label{disorderd}Critical dimension of disorder at the AA critical point}
In this section, we study the critical properties of disorder for $\delta=0$. It is expected that the localization length $\xi$ scales with the disorder strength $\Delta$ as $\xi\propto \Delta^{-\nu_\Delta}$ with $\nu_\Delta$ being the corresponding critical exponent. To determine $\nu_\Delta$, the finite-size scaling should be taken into account. Accordingly, scaling analysis gives the scaling form of $\xi$,
\begin{equation}
\label{Eq:xiscaling2}
\xi=Lf_1(\Delta L^{1/\nu_\Delta}),
\end{equation}
in which $f_i$ is the scaling function. When $L\rightarrow\infty$, Eq.~(\ref{Eq:xiscaling2}) recovers the scaling relation $\xi\propto \Delta^{-\nu_\Delta}$.

\begin{figure}[tbp]
\centering
  \includegraphics[width=0.8\linewidth,clip]{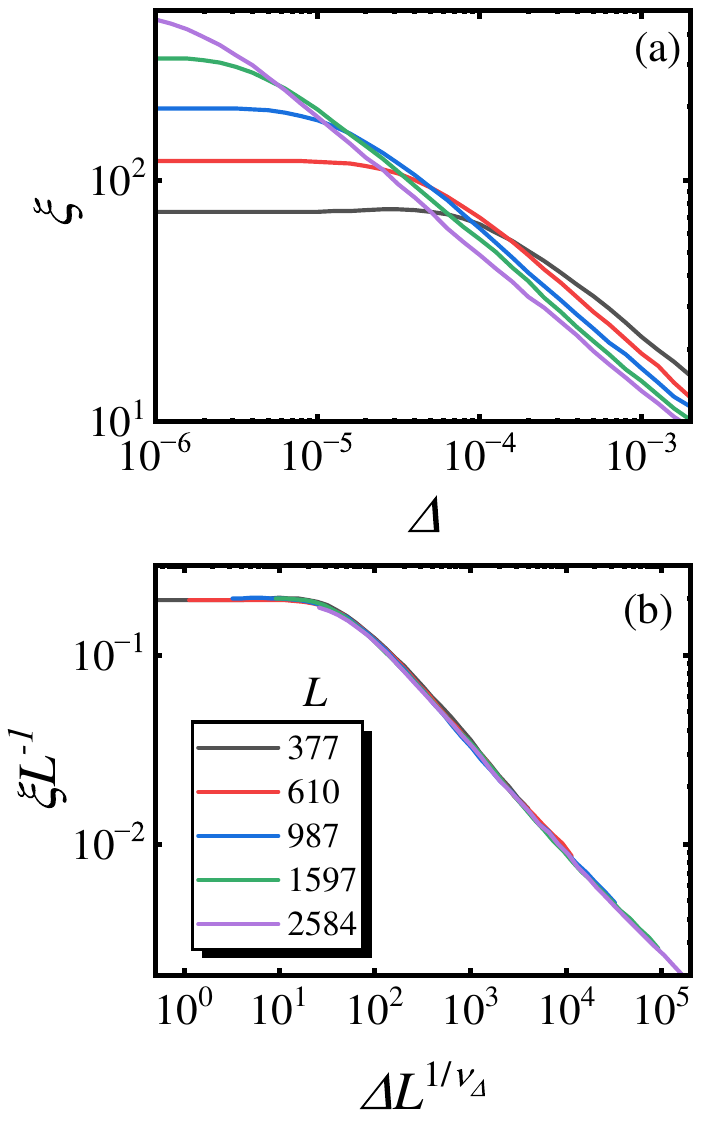}
  \vskip-3mm
  \caption{(a) Curves of $\xi$ in the ground state versus $\Delta$ for various $L$ at $\delta=0$. (b) Rescaled curves of $\xi L^{-1}$ versus $\Delta L^{-1/\nu_{\Delta}}$ collapse onto each other for $\nu_\Delta=0.46$. Double-log scales are used. The result is averaged for 1000 choices of $\phi$ and disorder.
  }
  \label{aacdis}
\end{figure}

We calculate the curves of $\xi$ versus $\Delta$ for various $L$ and show the results in Fig.~\ref{aacdis}. We estimate $\nu_\Delta$ by rescaling $\xi$ and $\Delta$ as $\xi L^{-1}$ and $\Delta L^{1/\nu_\Delta}$, respectively, for some trial values of $\nu_\Delta$. We find that the curves collapse onto each other quite well when $\nu_\Delta=0.46$, as shown in Fig.~\ref{aacdis} (b).

\begin{figure}[tbp]
\centering
  \includegraphics[width=0.8\linewidth,clip]{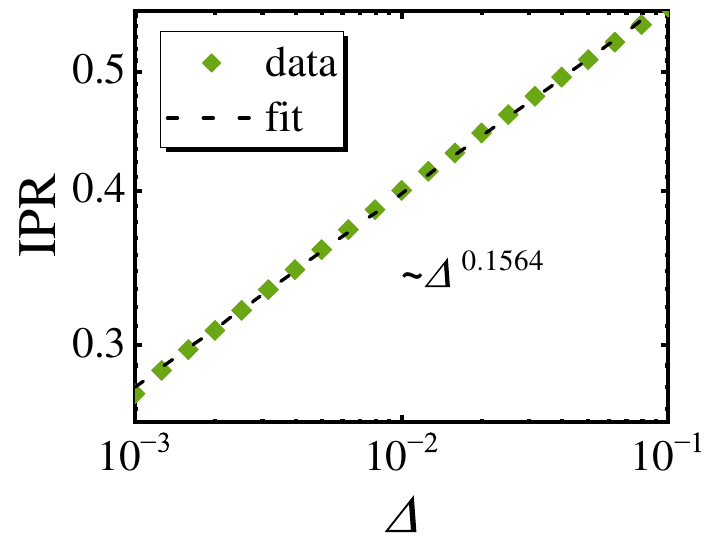}
  \vskip-3mm
  \caption{Curves of ${\rm IPR}$ in the ground state versus $\Delta$ at $\delta=0$ for $L=2584$. Power fit shows ${\rm IPR}\propto\Delta^{0.1564}$.  The result is averaged for 1000 choices of $\phi$ and disorder. Double-log scales are used.
  }
  \label{aac2584IPRs}
\end{figure}

To verify the value of $\nu_\Delta$, we explore the scaling property of ${\rm IPR}$ for $\delta=0$. Scaling analyses by setting $g=\delta$ and $s=s_\delta$ in Eq.~(\ref{Eq:iprscaling2}) give the scaling relation in presence of the disorder,
\begin{equation}
\label{Eq:iprscaling4}
{\rm IPR}=\delta^{s_\delta} f_2(\Delta \delta^{-\nu_\delta/\nu_\Delta}),
\end{equation}
in the thermodynamic limit. When $\delta\rightarrow 0$, Eq.~(\ref{Eq:iprscaling4}) gives
\begin{equation}
\label{Eq:iprscaling5}
{\rm IPR}\propto \Delta^{s_\delta\nu_\Delta/\nu_\delta},
\end{equation}
in which $s_\delta\nu_\Delta/\nu_\delta\approx0.153$ by setting $\nu_\Delta=0.46$ as input.
We calculate ${\rm IPR}$ versus $\Delta$ at $\delta=0$ for a large enough lattice size so that the size effects are tiny.
The result is shown in Fig.~\ref{aac2584IPRs}.
One finds that for $L=2584$ and the parameter region shown in Fig.~\ref{aac2584IPRs}, ${\rm IPR}\propto\Delta^{0.1564}$ with the exponent very close to $0.153$, confirming the value of $\nu_\Delta$.

Apparently, $\nu_\Delta$ is a new critical exponent since it is apparently different from both $\nu_\delta$ and $\nu_A$, demonstrating that the disorder contributes a new relevant direction in the AA critical point.
Furthermore, one finds that $\nu_\Delta$ is smaller than $\nu_\delta$, indicating that the disorder is less relevant than the quasiperiodic potential near the AA critical point.
A possible reason is that the disorder is short-range correlated but the quasiperiodic potential is long-range correlated.

\section{\label{scalingcr}Scaling properties in the critical region of the disordered AA model}
In this section, we explore the scaling properties in the critical region of the disordered AA model.

\subsection{\label{scalingf}General scaling forms in the critical region}
As shown in Fig.~\ref{phase}, a critical region near the AA critical point is spanned by the quasiperiodic potential $\delta$ and the disorder strength $\Delta$.
In the following, we show that scaling behaviors of the characteristic quantities, introduced in Sec.~\ref{secmodel}, can be described by the scaling forms with $\delta$ and $\Delta$ as the scaling variables.
In particular, for $\delta<0$, as a result of the presence of the overlapping regions between the disordered AA critical region and the Anderson transition region, a constraint should be imposed on the scaling functions~\cite{zhai2018}, giving a hybrid scaling form.

\begin{figure}[tbp]
\centering
  \includegraphics[width=\linewidth,clip]{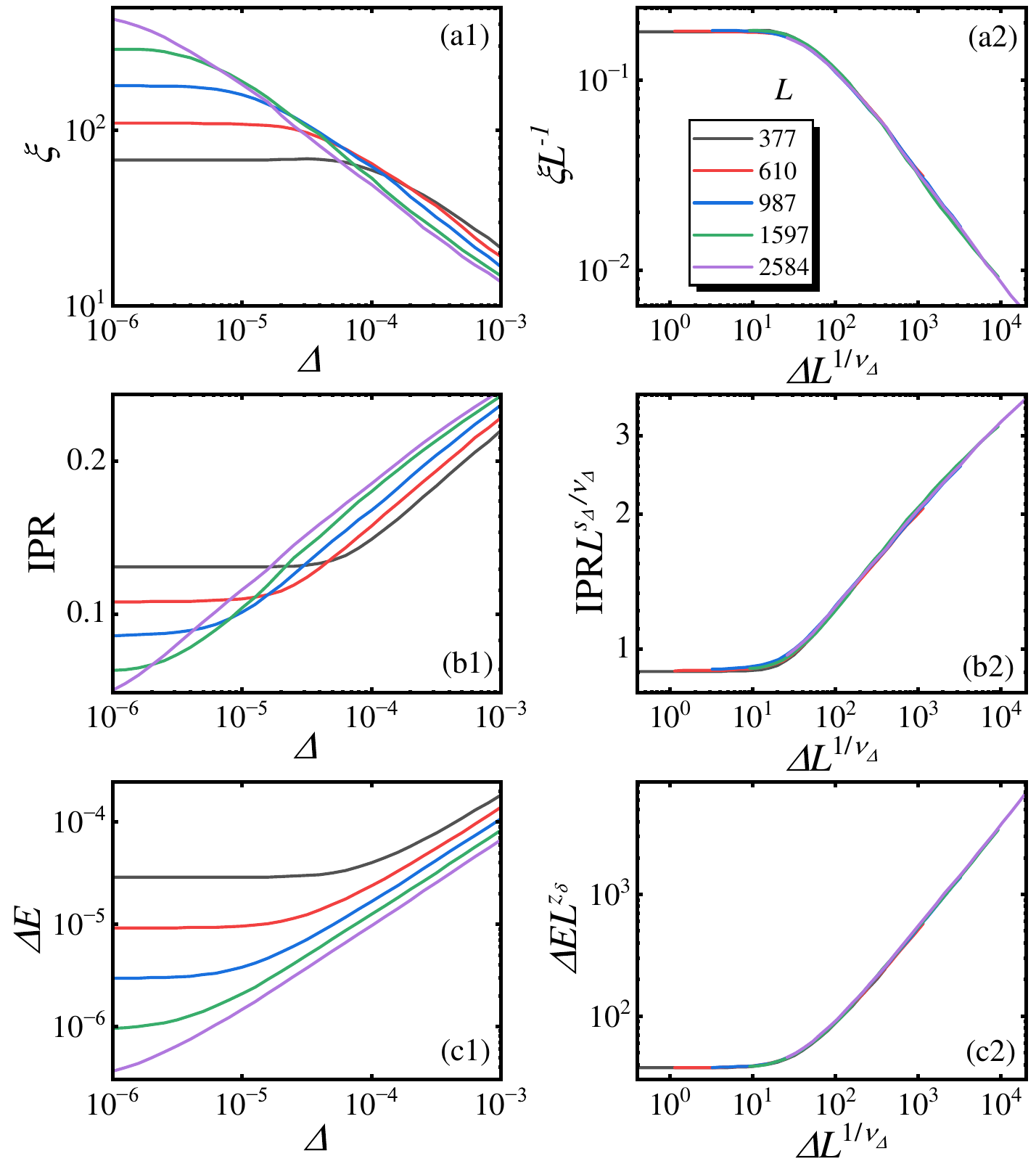}
  \vskip-3mm
  \caption{Scaling properties in the ground state for fixed $\delta L^{1/\nu_\delta}=0.1$. The curves of $\xi$ versus $\Delta$ before (a1) and after (a2) rescaled for different $L$.
  The curves of ${\rm IPR}$ versus $\Delta$ before (b1) and after (b2) rescaled for different $L$. And the curves of $\Delta E$ versus $\Delta$ before
  (c1) and after (c2) rescaled for different $L$.
  The result is averaged for 1000 choices of $\phi$ and disorder.}
  \label{deltala0re}
\end{figure}
\begin{figure}[tbp]
\centering
  \includegraphics[width=\linewidth,clip]{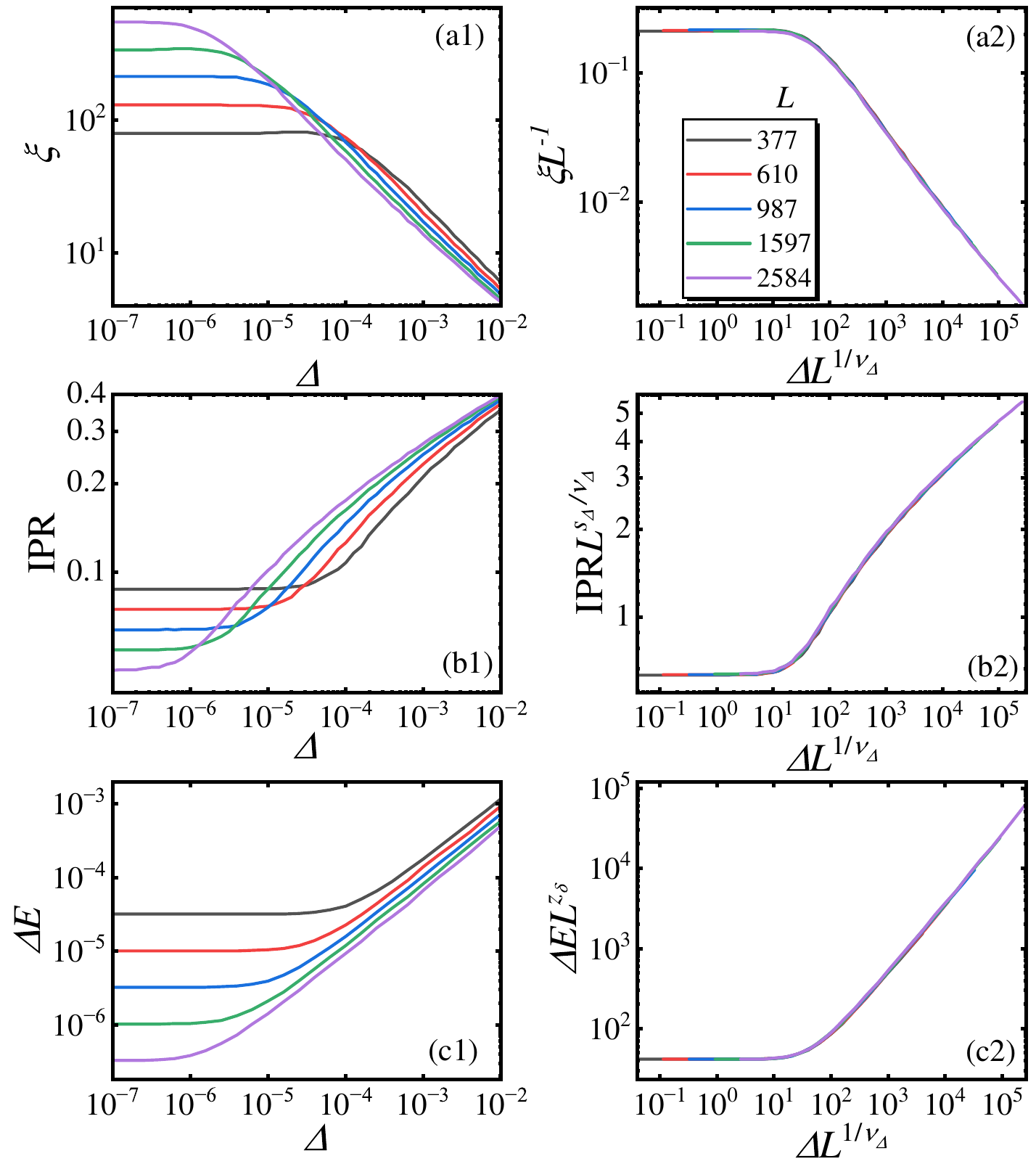}
  \vskip-3mm
  \caption{Scaling properties in the ground state for fixed $\delta L^{1/\nu_\delta}=-0.1$. The curves of $\xi$ versus $\Delta$ before (a1) and after (a2) rescaled for different $L$. The curves of IPR versus $\Delta$ before (b1) and after (b2) rescaled for different $L$.
  The curves of $\Delta E$ versus $\Delta$ before (c1) and after (c2) rescaled for different $L$. The result is averaged for 5000 choices of $\phi$ and disorder.
  }
  \label{deltasm0re}
\end{figure}

For general $\delta$ and $\Delta$ near the critical point, the localization length, $\xi$, should obey the scaling form
\begin{equation}
\label{Eq:xiscaling3}
\xi=Lf_3(\delta L^{1/\nu_\delta},\Delta L^{1/\nu_\Delta}).
\end{equation}
For $\Delta=0$ and $L\rightarrow\infty$, Eq.~(\ref{Eq:xiscaling3}) recovers Eq.~(\ref{Eq:xiscaling1}) in which $g=\delta$ and $\nu=\nu_\delta$; while for $g=0$, Eq.~(\ref{Eq:xiscaling3}) recovers Eq.~(\ref{Eq:xiscaling2}) which has been employed to estimate $\nu_\Delta$.

Similar full scaling forms can also be obtained for ${\rm IPR}$ and $\Delta E$. The scaling form for ${\rm IPR}$ should satisfy
\begin{equation}
\label{Eq:iprscaling6}
{\rm IPR}=L^{-s_\delta/\nu_\delta} f_4(\delta L^{1/\nu_\delta},\Delta L^{1/\nu_\Delta}).
\end{equation}
When $\Delta=0$ and $L\rightarrow\infty$, Eq.~(\ref{Eq:iprscaling6}) restores Eq.~(\ref{Eq:iprscaling1}) in which $g=\delta$ and $s=s_\delta$; when $\Delta=0$ and $g=0$, Eq.~(\ref{Eq:iprscaling6}) recovers Eq.~(\ref{Eq:iprscaling1}) in which $g=\delta$, $\nu=\nu_\delta$ and $s=s_\delta$; when $L\rightarrow\infty$, Eq.~(\ref{Eq:iprscaling6}) recovers Eqs.~(\ref{Eq:iprscaling4}) and (\ref{Eq:iprscaling5}), which have been used to verify the value of $\nu_\Delta$. In addition, the scaling form of $\Delta E$ should be
\begin{equation}
\label{Eq:gapscaling3}
\Delta E=L^{-z_\delta} f_5(\delta L^{1/\nu_\delta},\Delta L^{1/\nu_\Delta}).
\end{equation}
When $\Delta=0$ and $\delta=0$, Eq.~(\ref{Eq:gapscaling3}) recovers Eq.~(\ref{Eq:gapscaling}) in which $z=z_\delta$; when $\Delta=0$ and $L\rightarrow \infty$, Eq.~(\ref{Eq:gapscaling3}) recovers Eq.~(\ref{Eq:gapscaling2}) in which $z=z_\delta$ and $\nu=\nu_\delta$. These scaling forms Eqs.~(\ref{Eq:xiscaling3})-(\ref{Eq:gapscaling3}) should be applicable in the critical region surrounding the AA critical point, as shown in Fig.~\ref{phase}.

Moreover, in the region with $\delta<0$, the critical region of the Anderson localization also play significant roles. We take the scaling property of $\xi$ as the example. From the viewpoint of the Anderson transition, for a fixed $\delta$, $\xi$ should satisfy
\begin{equation}
\label{Eq:xiscaling4}
\xi=Lf_6(\Delta L^{1/\nu_A}).
\end{equation}
Thus, in the overlapping region between the AA critical region and the Anderson critical region, as illustrated in Fig.~\ref{phase}, $\xi$ should simultaneously obey Eqs.~(\ref{Eq:xiscaling3}) and ($\ref{Eq:xiscaling4}$).
This condition dictates that the scaling function of $f_3$ should satisfy the following hybrid scaling form for $\delta<0$,
\begin{equation}
\label{Eq:overlapscaling}
f_3(\delta L^{1/\nu_\delta},\Delta L^{1/\nu_\Delta})=f_7[\Delta L^{1/\nu_\Delta}(\delta L^{1/\nu_\delta})^\kappa],
\end{equation}
in which
\begin{equation}
\label{Eq:overlapscaling1}
\kappa\equiv\nu_\delta(1/\nu_A-1/\nu_\Delta),
\end{equation}
includes the critical exponents from both the AA model and the Anderson transition. Accordingly, the scaling function $f_7$ in Eq.~(\ref{Eq:overlapscaling}) provides the link between $f_3$ and $f_6$.

\begin{figure}[tbp]
\centering
  \includegraphics[width=0.9\linewidth,clip]{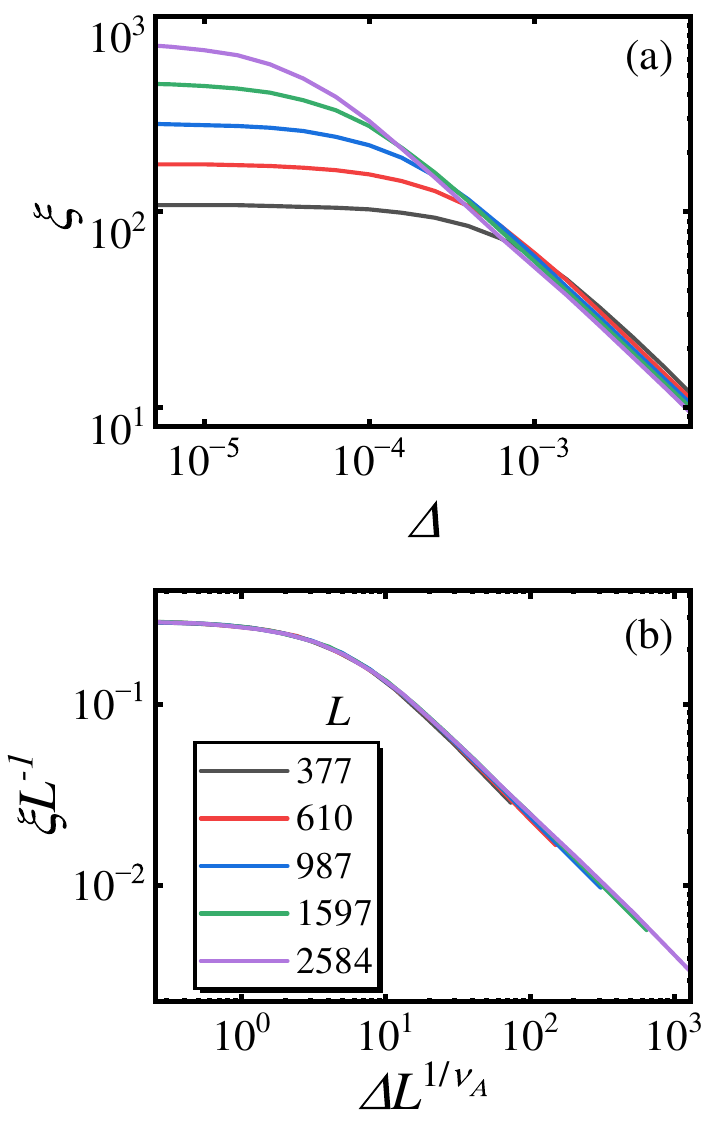}
  \vskip-3mm
  \caption{(a) Curves of $\xi$ versus $\Delta$ for various $L$ for fixed $\delta=-0.1$. (b) Rescaled curves of $\xi L^{-1}$ versus $\Delta L^{-1/\nu_{A}}$ collapse onto each other. Double-log scales are used. The result is averaged for 5000 choices of $\phi$ and disorder.
  }
  \label{pureAnderson}
\end{figure}

\begin{figure}[tbp]
\centering
  \includegraphics[width=0.9\linewidth,clip]{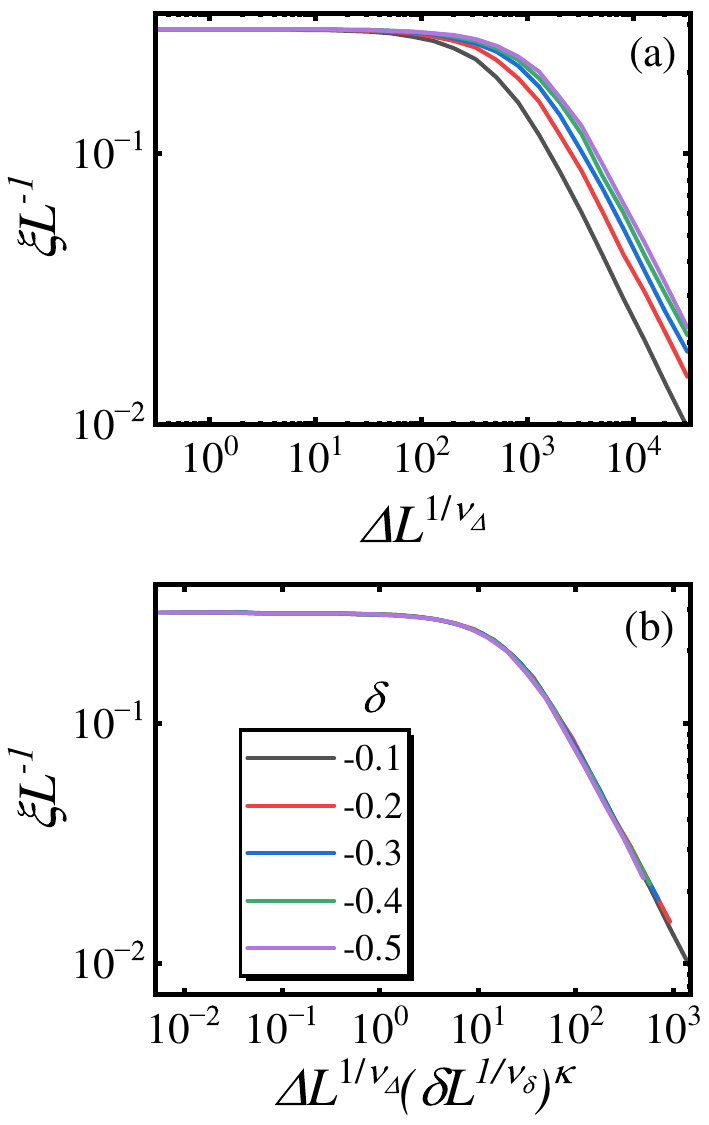}
  \vskip-3mm
  \caption{(a) Curves of $\xi L^{-1}$ versus $\Delta L^{-1/\nu_{\Delta}}$ for different choices of $\delta$. (b) Rescaled curves of $\xi L^{-1}$ versus $\Delta L^{1/\nu_\Delta}(\delta L^{1/\nu_\delta})^\kappa$ collapse onto each other. $L=987$, Double-log scales are used. The result is averaged for 5000 choices of $\phi$ and disorder.}
  \label{fhybrid}
\end{figure}

\subsection{Numerical results}
Firstly, we verify Eqs.~(\ref{Eq:xiscaling3})-(\ref{Eq:gapscaling3}). As shown in Figs.~\ref{deltala0re} and \ref{deltasm0re}, with fixed $\delta L^{1/\nu_\delta}$, we calculate curves of $\xi$, ${\rm IRP}$, and $\Delta E$ versus $\Delta$ for different lattice sizes. In Fig.~\ref{deltala0re}, we show the results for $\delta>0$, and in Fig.~\ref{deltasm0re}, we show the results for $\delta<0$. For both cases, we find that the rescaled curves collapse onto each other very well. These results not only demonstrate that the critical properties of the disordered AA model~(\ref{Eq:model}) can be described by the scaling theory including both $\delta$ and $\Delta$ as its relevant scaling variables, but also verify that the scaling dimension of $\Delta$ is $\nu_\Delta\approx 0.46$.

Secondly, as shown in Fig.~\ref{pureAnderson}, for a fixed $\delta<0$, which is still near the AA critical point, we calculate the curves of $\xi$ versus $\Delta$ for various $L$. After rescaling $\xi$ as $\xi L^{-1}$ and rescaling $\Delta$ as $\Delta L^{1/\nu_A}$ with $\nu_A=2/3$ being the critical exponent for the Anderson localization transition, we find that the rescaled curves match with each other perfectly. This result demonstrates that the scaling form of the Anderson localization Eq.~(\ref{Eq:xiscaling4}) is still applicable in the critical region of the disordered AA model. Combining these results with the previous ones, one confirms that the scaling behaviors for $\delta<0$ can be simultaneously described by both the critical theory of the disordered AA model and the theory of the Anderson localization.

Thirdly, we focus on the property of scaling function $f_3$ for $\delta<0$. We calculate the curves of $f_3=\xi L^{-1}$ versus $\Delta L^{1/\nu_\Delta}$ for various $\delta$ with a fixed $L$. We find that by rescaling $\Delta L^{1/\nu_\Delta}$ as $\Delta L^{1/\nu_\Delta}(\delta L^{1/\nu_\delta})^\kappa$, the curves collapse onto a single one, demonstrating that although two scaling variables are included in $f_3$, they are not independent ones in the overlapping critical region, as a result of the constraint given by the hybrid scaling form given by Eqs.~(\ref{Eq:overlapscaling}) and (\ref{Eq:overlapscaling1}).

\section{\label{secSum}Summary}
In summary, we have studied the quantum criticality of the disordered AA model. We have discovered that the disorder provides a new independent relevant direction in the AA critical point with the scaling dimension $\nu_\Delta\approx0.46$. Critical properties in the critical region spanned by the disorder strength and the quasiperiodic potential strength have been systematically explored. In particular, for $\delta<0$, there is an overlapping region between the AA critical region and the Anderson localization critical region. We have shown that the scaling behaviors satisfy both the critical theory of the AA critical point and the critical theory of the Anderson transition, giving a hybrid scaling form on the scaling function.

Our present work uncovers interesting critical properties in the localization transition. As possible generalizations, one can also investigate the many-body localization transition behavior simultaneously in presence of the quasiperiodic potential and the disorder. Note that there are some relevant works in this issue with some controversial problems~\cite{Khemani2017,zhangsx2018,Zhang2019arxiv}.
In addition, in contrast to the Hermitian case in which localization happens for infinitesimal disorder strength, in the non-Hermitian systems, localization happens for finite disorder strength ~\cite{Hatano1996,Goldsheid1998}.
Thus, the investigation on effects induced by disorder in the non-Hermitian AA model is intriguing and this work is in process.

\section*{Acknowledgments}
We would like to thank Shi-Xin Zhang for his helpful discussion. X.B. and S.Y. is supported by the Science and Technology Projects in Guangzhou (202102020367) and the Fundamental Research Funds for Central Universities (22qntd3005).
L.-J.Z. is supported by the National Natural Science Foundation of China (Grant No. 12274184) and China
Postdoctoral Science Foundation (Grant No. 2021M691535).


\begin{thebibliography}{99}
\bibitem{Harper1955}P. G. Harper, Proc. Phys. Soc. A {\bf68}, 874 (1955).
\bibitem{Aubry1980}S. Aubry and G. Andre, Ann. Israel Phys. Soc. {\bf3}, 133 (1980).
\bibitem{Sokoloff1981}J. B. Sokoloff, Phys. Rev. B {\bf23}, 6422 (1981).
\bibitem{Sarma1988}S. Das Sarma, S. He, and X. C. Xie, Phys. Rev. Lett. {\bf61}, 2144 (1988).
\bibitem{Biddle2009}J. Biddle, B. Wang, D. J. Priour, and S. Das Sarma, Phys. Rev. A {\bf80}, 021603 (2009).
\bibitem{Biddle2011}J. Biddle, D. J. Priour, B. Wang, and S. Das Sarma, Phys. Rev. B {\bf83}, 075105 (2011).
\bibitem{Lang2012}L.-J. Lang, X. Cai, and S. Chen, Phys. Rev. Lett. {\bf108}, 220401 (2012).
\bibitem{verbin2013}M. Verbin, O. Zilberberg, Y. E. Kraus, Y. Lahini, and Y. Silberberg, Phys. Rev. Lett. {\bf110}, 076403 (2013).
\bibitem{Zhang2018}D.-W. Zhang, Y.-Q. Zhu, Y. X. Zhao, H. Yan, and S.-L. Zhu, Adv. Phys. {\bf67}, 253 (2018).
\bibitem{Ganeshan2013}S. Ganeshan, K. Sun, and S. Das Sarma, Phys. Rev. Lett. {\bf110}, 180403 (2013).
\bibitem{Luschen2018}H. P. L\"uschen, S. Scherg, T. Kohlert, M. Schreiber, P. Bordia, X. Li, S. Das Sarma, and I. Bloch, Phys. Rev. Lett. {\bf120}, 160404 (2018).
\bibitem{Skipetrov2018} S. E. Skipetrov and A. Sinha, Phys. Rev. B {\bf97}, 104202 (2018).
\bibitem{Agrawal2020}U. Agrawal, S. Gopalakrishnan, and R. Vasseur, Nat. Commun. {\bf11}, 2225 (2020).
\bibitem{Longhi2019}S. Longhi, Phys. Rev. Lett. {\bf122}, 237601 (2019).
\bibitem{HepengYao2019}H. Yao, A. Khoudli, L. Bresque, and L. Sanchez-Palencia, Phys. Rev. Lett. {\bf123}, 070405 (2019).
\bibitem{zhai2020}L.-J. Zhai, S. Yin, and G.-Y. Huang, Phys. Rev. B {\bf102}, 064206 (2020).
\bibitem{Goblot2020}V. Goblot, A. \v{S}trkalj, N. Pernet, J. L. Lado, C. Dorow, A. Lema\^{i}tre, L. Le Gratiet, A. Harouri, I. Sagnes, S. Ravets, A. Amo, J. Bloch, O. Zilberberg, J. Bloch, and O. Zilberberg, Nat. Phys. {\bf16}, 832 (2020).
\bibitem{YFWang2022}Y. Wang, J.-H. Zhang, Y. Li, J. Wu, W. Liu, F. Mei, Y. Hu, L. Xiao, J. Ma, C. Chin, and S. Jia, Phys. Rev. Lett. {\bf129}, 103401 (2022).
\bibitem{CYuan2018}Y. Cao, V. Fatemi, S. Fang, K. Watanabe, T. Taniguchi, E. Kaxiras, and P. Jarillo-Herrero, Nature {\bf556}, 43 (2018).
\bibitem{Kraus2012}Y. E. Kraus, Y. Lahini, Z. Ringel, M. Verbin, and O. Zilberberg, Phys. Rev. Lett. {\bf109}, 106402 (2012).
\bibitem{Andrade2015}E. C. Andrade, A. Jagannathan, E. Miranda, M. Vojta, and V. Dobrosavljevi\'{c}, Phys. Rev. Lett. {\bf115}, 036403 (2015).
\bibitem{Aulbach2004}C. Aulbach, A. Wobst, G.-L. Ingold, P. H\"{a}nggi, and I. Varga, New J. Phys. {\bf6}, 70 (2004).
\bibitem{Billy2008}J. Billy, V. Josse, Z. Zuo, A. Bernard, B. Hambrecht, P. Lugan, D. Clment, L. Sanchez-Palencia, P. Bouyer, and A. Aspect, Nature {\bf453}, 891 (2008).
\bibitem{Roati2008}G. Roati, C. D'Errico, L. Fallani, M. Fattori, C. Fort, M. Zaccanti, G. Modugno, M. Modugno, and M. Inguscio, Nature {\bf453}, 895 (2008).
\bibitem{Liu2015}F. Liu, S. Ghosh, and Y. D. Chong, Phys. Rev. B {\bf91}, 014108 (2015).
\bibitem{Thouless1983}D. J. Thouless, Phys. Rev. B {\bf28}, 4272 (1983).
\bibitem{Wang2021}Y. Wang, C. Cheng, X.-J. Liu, and D. Yu, Phys. Rev. Lett. {\bf126}, 080602 (2021).
\bibitem{Strkalj2021}A. \v{S}trkalj, E. V. H. Doggen, I. V. Gornyi, and O. Zilberberg, Phys. Rev. Research {\bf3}, 033257 (2021).
\bibitem{Xu2019}S. Xu, X. Li, Y.-T. Hsu, B. Swingle, and S. Das Sarma, Phys. Rev. Research {\bf1}, 032039 (2019).
\bibitem{Mastropietro2015}V. Mastropietro, Phys. Rev. Lett. {\bf115}, 180401 (2015).
\bibitem{Khemani2017}V. Khemani, D. N. Sheng, and D. A. Huse, Phys. Rev. Lett. {\bf119}, 075702 (2017).
\bibitem{zhangsx2018}S.-X. Zhang and H. Yao, Phys. Rev. Lett. {\bf121}, 206601 (2018).
\bibitem{Zhang2019arxiv}S.-X. Zhang and H. Yao, arXiv, 1906.00971 (2019).
\bibitem{Hamizaki2019}R. Hamazaki, K. Kawabata, and M. Ueda, Phys. Rev. Lett. {\bf123}, 090603 (2019).
\bibitem{Jiang2019}H. Jiang, L.-J. Lang, C. Yang, S.-L. Zhu, and S. Chen, Phys. Rev. B {\bf100}, 054301 (2019).
\bibitem{Liu20212}Y. Liu, Q. Zhou, and S. Chen, Phys. Rev. B {\bf104}, 024201 (2021).
\bibitem{Schiffer2021}S. Schiffer, X.-J. Liu, H. Hu, and J. Wang, Phys. Rev. A {\bf103}, L011302 (2021).
\bibitem{Zhai2022}L.-J. Zhai, G.-Y. Huang, and S. Yin, Phys. Rev. B {\bf106}, 014204 (2022).
\bibitem{Evers2008}F. Evers and A. D. Mirlin, Rev. Mod. Phys. {\bf80}, 1355 (2008).

\bibitem{Anderson1958}P. W. Anderson, Phys. Rev. {\bf109}, 1492 (1958).
\bibitem{Thouless1974}D. J. Thouless, Phys. Rep. {\bf13}, 93 (1974).
\bibitem{Yang2021}H. Yang, J. Zeng, Y. Han, and Z. Qiao, Phys. Rev. B {\bf104}, 115414 (2021).
\bibitem{Pu2022}S. Pu, G. J. Sreejith, and J. K. Jain, Phys. Rev. Lett. {\bf128}, 116801 (2022).
\bibitem{Shinobu1981}S. Hikami, Phys. Rev. B {\bf24}, 2671 (1981).
\bibitem{Alexander1997}A. Altland and M. R. Zirnbauer, Phys. Rev. B {\bf55}, 1142 (1997).
\bibitem{TWang2021}T. Wang, T. Ohtsuki, and R. Shindou, Phys. Rev. B {\bf104}, 014206 (2021).
\bibitem{XLuo2021}X. Luo, T. Ohtsuki, and R. Shindou, Phys. Rev. Lett. {\bf126}, 090402 (2021).
\bibitem{XLuo2022}X. Luo, Z. Xiao, K. Kawabata, T. Ohtsuki, and R. Shindou, Phys. Rev. Research {\bf4}, L022035 (2022).
\bibitem{Semeghini2015}G. Semeghini, M. Landini, P. Castilho, S. Roy, G. Spagnolli, A. Trenkwalder, M. Fattori, M. Inguscio, and G. Modugno, Nat. Phys. {\bf11}, 554 (2015).
\bibitem{DHWhite2020}D. H. White, T. A. Haase, D. J. Brown, M. D. Hoogerland, M. S. Najafabadi, J. L. Helm, C. Gies, D. Schumayer, and D. A. W. Hutchinson, Nat. Commun. {\bf11}, 4942 (2020).
\bibitem{Wiersma1997}D. S. Wiersma, P. Bartolini, A. Lagendijk, and R. Righini, Nature {\bf390}, 671 (1997).
\bibitem{Mookherjea2014}S. Mookherjea, J. R. Ong, X. Luo, and L. Guo-Qiang, Nature Nanotech. {\bf9}, 365 (2014).
\bibitem{HFHu2008}H. Hu, A. Strybulevych, J. H. Page, S. E. Skipetrov, and B. A. van Tiggelen, Nat. Phys. {\bf4}, 945 (2008).
\bibitem{Lee1985}P. A. Lee and T. V. Ramakrishnan, Rev. Mod. Phys. {\bf57}, 287 (1985).
\bibitem{Subirbook}S. Sachdev, \emph{Quantum Phase Transitions}, 2nd ed. (Cambridge University Press, 2011).
\bibitem{MatthiasVojta2003}M. Vojta, Rep. Prog. Phys. {\bf66}, 2069 (2003).
\bibitem{Coleman2005}P. Coleman and A. J. Schofield, Nature {\bf433}, 226 (2005).
\bibitem{Sinha2019}A. Sinha, M. M. Rams, and J. Dziarmaga, Phys. Rev. B {\bf99}, 094203 (2019).
\bibitem{Wei2019}B.-B. Wei, Phys. Rev. A {\bf99}, 042117 (2019).
\bibitem{Bauer1990} J. Bauer, T. M. Chang, and J. L. Skinner, Phys. Rev. B {\bf42}, 8121 (1990).
\bibitem{Fyodorov1992}Y. V. Fyodorov and A. D. Mirlin, Phys. Rev. Lett. {\bf69}, 1093 (1992).
\bibitem{Cestari2011}J. C. C. Cestari, A. Foerster, M. A. Gusm\~{a}o, and M. Continentino, Phys. Rev. A {\bf84}, 055601 (2011).
\bibitem{zhai2018}L.-J. Zhai, H.-Y. Wang, and S. Yin, Phys. Rev. B {\bf97}, 134108 (2018).
\bibitem{Hatano1996}N. Hatano and D. R. Nelson, Phys. Rev. Lett. {\bf77}, 570 (1996).
\bibitem{Goldsheid1998}I. Y. Goldsheid and B. A. Khoruzhenko, Phys. Rev. Lett. {\bf80}, 2897 (1998).

\end{thebibliography}
\end{document}